\documentclass[fleqn,twoside]{article}
\usepackage{espcrc2}

\usepackage{graphicx}

\usepackage{epsfig}

\usepackage[figuresright]{rotating}

\newcommand{\AmS}{{\protect\the\textfont2
  A\kern-.1667em\lower.5ex\hbox{M}\kern-.125emS}}

\title{Future physics with polarized ep colliders}

\author{A. De Roeck\address{CERN, 1211 Geneva 23, Switzerland}
\thanks{Invited talk presented at the
``International Workshop on the Spin Structure of the Proton and
Polarized Collider Physics'', ECT$^*$, Trento, Italy, 
July 23-28, 2001.}
}

\begin{document}

\begin{abstract}
We discuss some of the physics opportunities at polarized
$ep$-colliders. Recent studies for polarized $ep$-colliders include
HERA, EIC and THERA.
\vspace{1pc}
\end{abstract}

\maketitle

\section{Introduction}
\vspace{1mm}
\noindent

The commissioning of the first electron-proton collider HERA 
(27.5~GeV electrons on 820~GeV protons)
eight
years ago opened up a completely new kinematical domain 
in deep inelastic scattering (DIS), and the two HERA experiments have provided 
a multitude of new insights into the structure of the 
proton and the photon since then. 
It is therefore natural to 
assume that the operation of $ep$-colliders with polarized proton and electron 
beams
will add vital new information to our picture of the spin structure of the 
nucleon. 
 Possible high energy $ep$-collider projects
presently under discussion are listed in Table~\ref{tab1}.

HERA has just been upgraded
to reach a higher luminosity. It has excellent detectors, and the 
electron beam is already polarized.
The polarization of the proton beam is technically more 
involved than for the electron beam, since protons do not 
polarize naturally in the HERA ring.
Hence beams from a source of polarized protons have to be
accelerated through the whole chain and the 
polarization has to be kept during the process.
The technical aspects of this project are elaborated in~\cite{barber}. 
Based on these studies, it seems realistic to assume that 
HERA could be operated with polarized electron and proton beams, each  
polarized to about 70\%, acumulating a luminosity of 
 500~pb$^{-1}$  when integrated over several years.

The Electron Ion Collider (EIC)~\cite{erhic}
 --if built at BNL-- will need a polarized 
electron accelerator, most likely a LINAC, added to the RHIC polarized
proton rings, and will  probably also need a dedicated
experiment. THERA~\cite{thera}
 will need TESLA to be built at DESY (or tangential
to the TEVATRON), polarized protons in the proton ring, and a new
detector.
The advantage of the EIC
 is its large reachable luminosity, imperative for 
polarized studies. At HERA the luminosity is  (just) enough for most 
topics but its advantage lies in its larger kinematical reach. THERA 
reaches even further in the  $x$ and $Q^2$ plane, but its
relatively low luminosity of 40 pb$^{-1}$/year 
 may be a handicap for many studies.

Most detailed studies for polarized $ep$-scattering at a collider have been
made in a series of workshops for HERA~\cite{gehr,works} and these will
serve as a basis 
for the discussion in this paper. Many of these studies have been carried over
to EIC in the last two years. For THERA polarized studies have only been 
briefly explored so far.

Here we will discuss  measurements of  $g_1(x,Q^2)$ at low $x$, the 
extraction of the polarized gluon density $\Delta G(x,Q^2)$ in a wide 
kinematic range using several processes, the extraction of the 
spin structure function $g_5(x,Q^2)$ from charged current events,
photoproduction studies and studies on the helicity structure of the
high $Q^2$ region.

\begin{table}
\caption{Possible future $ep$-collider facilities for polarized scattering}
\label{tab1}
\begin{tabular}{|l|l|l|}
\hline
Machine  & Lumi/year  & $\sqrt{s}$ \\
\hline
HERA   & 150 pb$^{-1}$& 320 GeV \\
Electron-Ion & $ 4$ fb$^{-1}$ & 30-100 GeV\\
Collider (EIC) & & \\
THERA  & 40 (250) pb$^{-1}$ &  1-1.6 TeV \\
(TESLA$\otimes$HERA) & & \\
\hline
\end{tabular}
\end{table}

\section{The polarized structure function \boldmath $g_1(x,Q^2)$}
\begin{figure}[t!]
\begin{center}
\epsfig{file=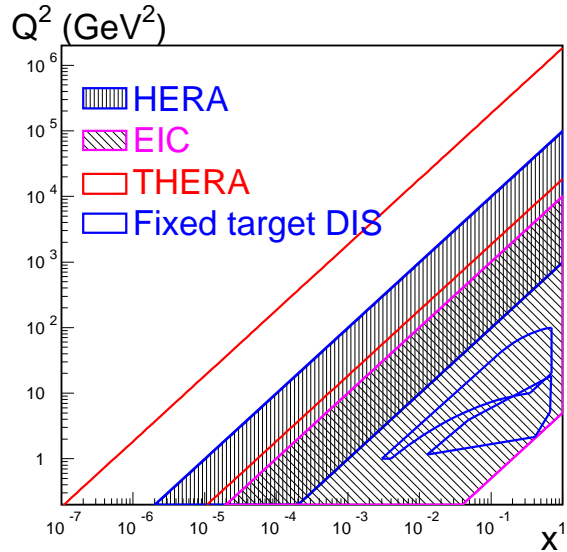,bbllx=90pt,bblly=230pt,bburx=500pt,bbury=600pt,width=8cm}
\caption{Measurable $x-Q^2$ region for a polarized HERA with the 
presently explored regions by fixed targer 
experiments. The energies used are for 
THERA: $E_p = 920, E_e=500$ GeV;
  HERA: $E_p = 920, E_e=27$ GeV;
  eRHIC: $E_p=250, E_e=10$ GeV and 
        $ E_p=50, E_e=5$ GeV, and roughly $0.01 < y < 1.0$.}
\label{fig:kin}
\end{center}
\end{figure}

\vspace{1mm}
\noindent
The outstanding advantage of a high energy $ep$-collider
 is that it can measure 
structure functions at very small $x$ 
and very large $Q^2$. The kinematical reach is shown in
Fig.~\ref{fig:kin} for the different
machines.
The region covered by present fixed target experiments is
shown as well. The region can be extended by several orders of magnitude 
both in $x$ and $Q^2$. 
The regions of the three colliders overlap.
At HERA measurements at 
 values of $Q^2$ up to 
$2\cdot 10^4$ GeV$^2$ are reached, and
values of $x$ below $  10^{-5}$. 
Hence such data  will allow for detailed QCD tests similar to the ones 
in the unpolarized case.

\begin{figure}[t!]
\begin{center}
~ \epsfig{file=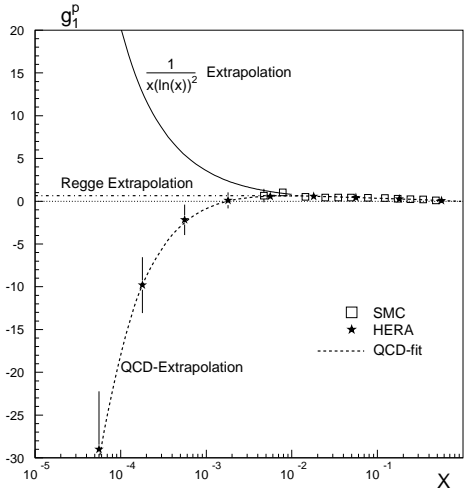,bbllx=0pt,bblly=0pt,bburx=560pt,bbury=560pt,width=7cm}
\caption{ The statistical uncertainty on the structure function $g_1$ of
the proton 
measurable at HERA, evolved to a value of 
 $Q^2 = 10$ GeV$^2$ for an integrated luminosity of
500 pb$^{-1}$. The SMC measurements are shown for comparison.}
\label{fig:g1p}
\end{center}
\end{figure}

An example for HERA is shown in Fig.~\ref{fig:g1p}.
The low-$x$ behaviour of $g_1$ is indicated in the figure:
the straight line is an 
extrapolation based on Regge phenomenology, and the upper curve
presents a scenario suggested in~\cite{abhay} where 
$g_1$ rises as $1/(x \ln^2(x))$, which is the maximally singular 
behaviour still consistent with integrability requirements.
The low $x$ behaviour of $g_1$ by itself is an interesting 
topic as discussed in~\cite{Kwiecins}.

\begin{figure}[t!]
\begin{center}
\epsfig{file=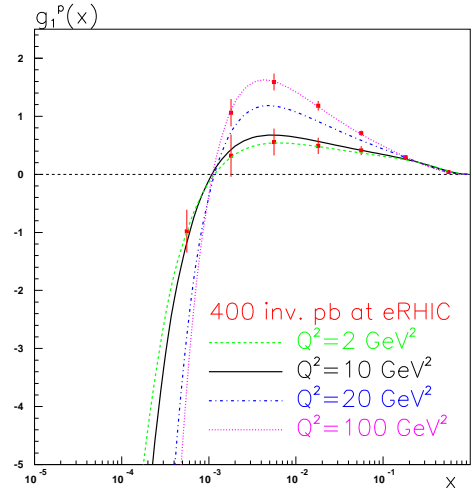,bbllx=0pt,bblly=40pt,bburx=550pt,bbury=550pt,width=7cm}
\caption{
The statistical uncertainty on the structure function $g_1$ of
the proton 
measurable at EIC,  
  for an integrated luminosity of
400 pb$^{-1}$.}
\label{fig:g1peic}
\end{center}
\end{figure}

An EIC has as expected a more restricted range, as shown in 
Fig.~\ref{fig:g1peic}. However the error bars can be substantially 
smaller due to the high luminosity of 4 fb$^{-1}$/year.
THERA would reach $x$ values down to 10$^{-6}$, but needs at least 
one fb$^{-1}$ of data.

Note that the expected asymmetries at low $x$, i.e. $x \sim 10^{-4}$ are 
relatively small, about $10^{-3}$, which puts strong requirements
on the control of the systematic effects, as discussed in~\cite{works}.
At the lowest $x$ values at THERA,  the asymmetries
may be  even smaller, and the potential to actually 
extract a polarized structure function at these values still needs to 
demonstrated.

It would be very advantageous to have 
polarized low-$x$ neutron data which would 
enable to measure
 singlet and non-singlet polarized structure functions at low $x$.
A study was made using polarized $^3$He at HERA~\cite{abhay}, from which 
$g_1^n$ can be extracted. If the machine can provide polarized $^3$He
with a luminosity comparable to the one for the protons, such a program
can be carried out.
A  recent new idea is to store polarized deutrons~\cite{derb}.
 Due to the small 
gyromagnetic anomaly
 value the storage and acceleration problems are less severe
for deutrons and
it could be possible to keep the polarization at HERA
even without the use of 
Siberian snake magnets for deuterons. However the spin
cannot
be flipped by spin rotators from transverse to longitudinal 
polarization in the interaction regions, and other means, such as the 
recently suggested use of external radial frequency fields must be considered
for arranging that the spins align longitudinally at the interaction 
points.
For polarized  deuteron beams
the changes to the present HERA machine could be more modest, and,
 if one can instrument the region around the beam-pipe to tag the 
spectator in the deuteron nucleus, it can give simultaneously
 samples of scattering on $p$ and $n$.
An intense polarized deuteron source is however needed. 
Furthermore, so far the  accelerating and storing
of polarized deuteron beams 
has been studied to a much lesser extend than 
for polarized protons.

With proton and neutron data available one can measure the Bjorken 
sum rule $\int g_1^p dx  -\int g_1^ndx$, 
presently measured to about 10\% precision.
E.g.  EIC data will allow to measure this sum rule, a key test for QCD, 
to a few \% precision.

\section{The polarized gluon distribution $\Delta G(x,Q^2)$}
\vspace{1mm}
\noindent
The high quality $g_1$ data from the fixed target experiments allows for 
quantitative QCD studies of the polarized structure function data, from 
which polarized parton distributions are extracted~\cite{abhay}. 
The relatively  small contribution of the quarks to
the spin of the proton,
as follows from the 'spin puzzle',
leads to a rather large value of the polarized gluon.
The precision is however rather limited.
 The measurements from present 
day data for the first moment of the 
polarized gluon distribution give typically $\int \Delta G(x)~{\rm d}x 
= 0.9 \pm 0.3({\rm exp})\pm 1.0({\rm theory})$ at $Q^2=1$~GeV$^2$.  
The theoretical error on this 
quantity is  dominated by the extrapolation into 
the yet unmeasured low-$x$ region.
Including 
future HERA data will improve the experimental error to about 0.2.
The theoretical error 
is expected to decrease by more than a 
factor 2 once $g_1(x,Q^2)$ is measured at low $x$.
At an EIC one expects to reduce the statistical uncertainty even to 0.08
with about 10 fb$^{-1}$, as discussed
in ~\cite{lichtenstadt}.

It is crucial for our full understanding of the proton spin that
the prediction of a large polarized gluon is 
 confirmed by direct measurements
of $\Delta G$.
HERA has shown
that the large centre of mass system (CMS) energy allows
for several processes to be used to extract the unpolarized gluon 
distribution.
These include jet 
and high $p_t$ hadron production,  charm production 
both in DIS and photoproduction, and correlations between 
multiplicities of the current and target hemisphere of the events
in the Breit frame. 


\begin{figure}[t!]
\begin{center}
\epsfig{file=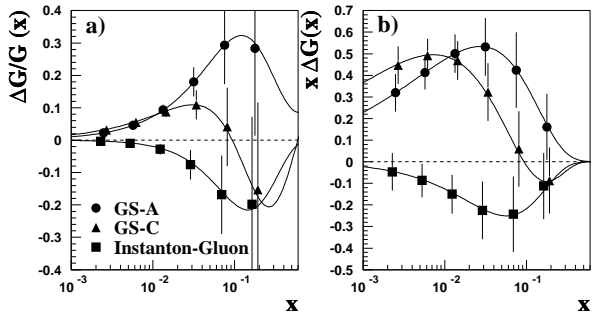,bbllx=30pt,bblly=275pt,bburx=580pt,bbury=565pt,width=8.5cm} 
\caption{High $p_t$ 
hadrons: Sensitivity to $\Delta G/G$ (a) and $x\Delta G$ (b)
 for  three different polarized gluon distributions shown
as solid lines and a luminosity
of 500~pb$^{-1}$, for $Q^2=20~{\rm GeV}^2$.}
\label{fig-t-500}
\end{center}
\end{figure}

The most promising process for  a direct 
extraction of  $\Delta G$ at HERA is   di-jet production, as discussed
in ~\cite{radel}.
The underlying idea is to isolate boson-gluon fusion events, i.e.~a process
where the gluon distribution enters at the Born level. 
 Jets are selected
with a $p_t>5$~GeV and are restricted to the acceptance 
of a typical existing HERA detector by 
the requirement $|\eta^{jet}_{LAB}| < 2.8$, where
$\eta^{jet}_{LAB}$ is the pseudo-rapidity in the laboratory system.
The resulting measurable range in  
$x$ (of the gluon) is $0.002 < x < 0.2$ with typically 6 measurable
 data points.
At an EIC
the measurable range is reduced to  
 $0.02 < x < 0.3$. 
The reach at THERA is  $0.0005 < x < 0.1$ but an event sample of 
several hundered  pb$^{-1}$
will be needed.

Instead of jets, single  hadrons 
with high transverse momentum $p_t$ opposite in azimuthal angle 
 in the $\gamma^*p$ frame can be used.~\cite{bravar}. 
In a study for HERA
 two charged tracks with a $p_t$ larger than 1.5 GeV are 
required~\cite{tracks}. 
The resulting gluon distribution is 
shown in Fig.~\ref{fig-t-500} and compared to 
several predicted gluon distributions. A similar level of discriminating power
as for the di-jet events is obtained, except in  the highest $x$ 
region, where the di-jets are  superior.
These measurements allow for the determination of the 
{\it shape} of $\Delta G(x)$. 
The errors on the individual points for the di-jet measurement
on $\Delta G(x)/G(x)$ are in the range from 0.007 to 0.1.
The total error on $\Delta G(x)/G(x)$ in the complete range is 0.02.

An exploratory study was made, using the values of $\Delta G(x)$
obtained from the di-jet 
analysis as an extra constraint in the fit of $g_1$ data discussed above.
The improvement of the errors on the first
moment of $\Delta G$  due to the inclusion of di-jet data
is shown in Table~\ref{tab-hera}.
\begin{table}[t]
\caption{\label{tab-hera} The expected statistical 
uncertainty in the determination of
  the first moment of the gluon distribution at $Q^{2} = 1$ GeV$^2$
(500~pb$^{-1}$), see text.}
\hfil
\begin{tabular}{||l|c||}
\hline\hline
  {\bf Analysis Type} &  {\boldmath $\delta (\int \Delta G~{\rm d}x)$} \\
\hline \hline
1. $g_1$ fixed target              &   0.3           \\
\hline
2.  $g_1$ fixed target + HERA   &   0.2           \\
\hline
3. di-jets at HERA                   &   0.2          \\
\hline
4. combined 2 \& 3                    &    0.1   \\
\hline\hline
\end{tabular}
\hfil
\end{table}
 The first two  rows give the values quoted before, namely for the NLO QCD
analysis without and with  projected
HERA data for $g_1$. The third row shows the expected error if
only the di-jet asymmetry is added to the fixed target 
 $g_1$ data, and the fourth
row shows the total improvement using all available information.


\begin{figure}[t!]
  \centering \epsfig{file=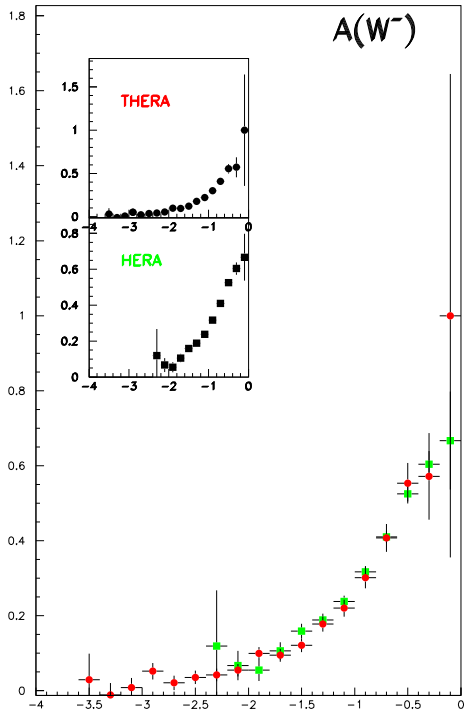,width=.38\textwidth}
\caption{Spin asymmetries $A^{W^-}$ 
   for charged current 
events for a total
  luminosity of 500 pb$^{-1}$. 
The error bars represent
  the statistical uncertainty of the measurement.} 
\label{fig_results}
\end{figure}

\section{Polarized quark distributions}
\label{sec:pquark}
\vspace{1mm}
\noindent

In present fixed target experiments information on the flavour decomposition 
can be obtained from  semi-inclusive measurements, 
i.e.~measurements where a final state hadron is tagged. 
At HERA one has the additional option to study quark flavours in 
a more inclusive, and therefore fragmentation function independent mode,
via charged current interactions.
 The asymmetry is defined by 
\begin{equation}
A^{W\mp} =
\frac{d\sigma^{W^\mp}_{\uparrow\downarrow}-d\sigma^{W^\mp}_{\uparrow\uparrow}}
{d\sigma^{W^\mp}_{\uparrow\downarrow}+d\sigma^{W^\mp}_{\uparrow\uparrow}}
\nonumber\\
\approx \frac{g^{W^{\mp}}_5}{F^{W^{\mp}}_1} \label{eq_as}
\end{equation}
with 
$g^{W^-}_5 = \Delta u+\Delta c - \Delta\bar{d} - 
\Delta
\bar{s} $, $g^{W^+}_5 = \Delta d+\Delta s - \Delta\bar{u} - \Delta\bar{c}$.
A Monte Carlo study, including detector effects, was made for the 
measurements of the asymmetry and the extraction of $g_5$~\cite{contreras2}.
The total missing transverse momentum (which is a signal for the 
escaping neutrino) was required to be $P_{Tmiss}>15$ GeV,  and the region 
$Q^2>225$ GeV$^2$ has been selected for this analysis. 
The results for the asymmetries for both HERA and THERA
 are shown in Fig.~\ref{fig_results}. 
The error bars indicate the statistical precision 
of the measurement. The asymmetries are very large
and significant measurements can be produced at THERA down to 
$x= 10^{-3}$.
These charged current measurements, with both 
$e^+$ and $e^-$ beams,   can be used to 
extract e.g.~the $\Delta u$ and $\Delta d$ distributions.

\section{Photoproduction}
\vspace{1mm}
\noindent
Photoproduction processes have been shown to be sensitive probes
for the polarized parton structure in the proton AND in the 
photon. Single jet production has been studied in~\cite{gehr}.
In~\cite{gamma} it was shown that  also 
a high $p_t$ track analyses yields 
 a similar sensitivity, with only modest luminosity 
requirements.

\begin{figure}[t!]
\begin{center}
~ \epsfig{file=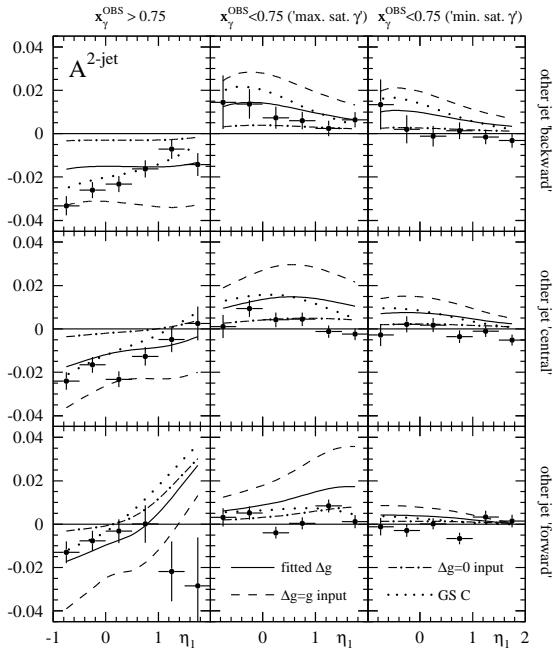,width=8cm}
\caption{Polarized photoproduction of di-jets ($E_{T,1} > 8$~GeV,
$E_{T,2} > 6$~GeV): 
asymmetries for direct (first column) and resolved (second and third 
column) photon contributions as function of the rapidity of the first 
jet and for different orientations of the second jet.
Second and third column correspond to different scenarios for 
the parton content of the polarized photon suggested 
in~\protect\cite{grvgamma}.
The error bars shown 
correspond to a Monte Carlo sample of 50~pb\protect{$^{-1}$}.
}
\label{fig:phoprod}
\end{center}
\end{figure}

Polarized photoproduction of di-jets has been investigated in detail,
 including in particular effects due to 
parton showering, hadronization, jet finding and jet clustering. 
It could be demonstrated that, although 
these effects yield sizable corrections, the measurable asymmetry will
 largely be preserved at the hadron level~\cite{gamma}. 
An example for the correspondence of parton and hadron level asymmetries is 
shown in Fig.~\ref{fig:phoprod}, obtained with a moderate integrated 
luminosity of only 50~pb$^{-1}$. 
A first idea on the discriminative power on the photon 
structure of future measurements 
can however be gained by comparing the predictions obtained 
with the two (minimal and maximal) polarization scenarios 
proposed in~\cite{grvgamma}, as done in Fig.~\ref{fig:phoprod}.

Finally, it can be pointed out  that a measurement of the total 
polarized photoproduction cross section $\Delta \sigma_{\gamma p} (\nu)$
as function of the photon-proton CMS energy $\nu$
at HERA would contribute to a precise understanding of the 
Drell-Hearn-Gerasimov sum rule, as discussed in~\cite{dhg}. 
In particular the  Regge behaviour of the energy dependence
of the polarized cross 
section can be tested. The sensitivity which could be reached with EIC is 
about one order of magnitude larger than the one for HERA.

\section{Effects at high $Q^2$}
\vspace{1mm}
\noindent

At an $ep$-collider the region of 
high $Q^2$ offers the largest chance of discovering new physics.
Such new physics could manifest itself through the production
of new particles, such as Leptoquarks or SUSY particles in RP
violating models, contact interactions, etc.
A general study was made based on the 
contact interaction formalism~\cite{virey}. 
 It was 
demonstrated that a fully polarized HERA would be very instrumental 
in disentangling the chiral structure of the new interactions.
With 250 pb$^{-1}$ data samples for  polarized $e^+$ and
$e^-$ beams, for each of the 2 spin orientations,
the asymmetries are sensitive to
contact interactions to scales
larger than 7 TeV (95\% C.L.).
In the presence of a signal these different
combinations of cross sections into the seven different asymmetries
allow a complete identification of the chiral structure of the new
interactions, i.e.~whether the interactions are LL, RR, LR or LR or 
a combination of those (where L and R denote the left and right handed 
fermion helicities  for the lepton and quark respectively).

This study has been further extended to the special case of 
leptoquark-like production~\cite{virey2}: asymmetries like the ones above
would allow to pin down the chiral properties of the couplings 
to these new
particles.

 An interesting 
possibility is  the effect induced by  QCD 
instantons~\cite{kochelev} to the proton structure function.
 Non-perturbative instanton fluctuations
describe the quantum tunnelling between different gauge rotated classical
vacua in QCD.
Due to the quark helicity flip at the quark-instanton vertex, the 
contribution to the spin-dependent cross sections of instantons is 
very different from the one of the perturbative quark-gluon vertex.
Furthermore, in the instanton liquid model~\cite{instanton} 
the  contribution of instantons to the  proton structure
is expected to  become increasingly more important with increasing
$Q^2$~\cite{kochelev}.  

These high $Q^2$ studies may be most relevant for HERA and THERA.
For THERA both $e^+$ and $e^-$ beams will be needed and
minimum integrated luminosities of 
order 100-200 pb$^{-1}$/beam.

\section{Other Topics}
Many other topics on polarization can be studied at an
$ep$-collider, e.g. 
target fragmentation properties, transversity, DHG, issues related
to diffraction, spin transfer in $\Lambda $ polarization,
semi-inclusive measurements to extract the polarized quark distributions,
etc. Further details can be found in~\cite{works}.

\section{Conclusions}

Polarized $ep$-colliders with a
 centre of mass energies of 100-1000 GeV
 will  allow to make unique 
measurements in polarized  deep inelastic scattering as well as
photoproduction.

The necessity
for low-$x$ measurements of the structure functions, and 
determination of the polarized gluon
distribution $\Delta G$ have 
been wildly advocated by the spin physics community over the last five years.
$ep$-colliders  can play a pivotal role in this field since they are able 
 to give conclusive insight on both of these issues.

Polarized $ep$-colliders 
 will also contribute to the flavour decomposition of the quark spin
distributions,  and the very intriguing possibility
to measure polarized parton distributions in the photon. 
Furthermore --not discussed in this review-- new insight is expected
on spin transfer in quark fragmentation and spin effects in 
diffractive scattering.
Finally, a polarized HERA or THERA will be very instrumental in the study and
interpretation of possible deviations from the Standard
Model expectation in the  high-$Q^2$
region, if observed.

However the required luminosity for any such machine should be at least
100 pb$^{-1}$/year, preferably even more. The polarizations of the beams
should be larger than 50-60\%. If EIC will be the world's only 
polarized collider, it would be useful to consider having an 
extention  of the CMS energy, via an upgrade of the 
$e$-LINAC energy e.g. from 10 to 20-30 GeV, to cover an as large
kinematic region as possible. If new detectors are required or 
can be afforded, particular care should be taken on the design, 
learning from shortcomings from the present HERA detectors, in
e.g.  the small angle instrumentation~\cite{krasny}.
Polarized deuterons seem an attractive addition or even alternative
to polarized protons, and are perhaps easier to accelerate and store, but 
detailed machine studies need to be performed to confirm these ideas.

In all, polarized scattering experiments have in the last decades
often revealed surprises, hence a polarized collider seems an attractive
future option 
with a  rich program.

\end{document}